\documentclass[manuscript]{aastex}

\def\ltsima{$\;\buildrel < \over \sim \;$}
\def\simlt{\lower.5ex \hbox{\ltsima}}
\def\gtsima{$\;\buildrel > \over \sim \;$}
\def\simgt{\lower.5ex \hbox{\gtsima}}
\begin{document}
\title{The Infrared Continuum Spectrum of VY CMa}

\author{Martin Harwit\altaffilmark{1,3}} 
\affil{511 H Street S.W., Washington, DC 20024--2725}

\author{Koen Malfait\altaffilmark{2}, Leen Decin, Christoffel Waelkens}
\affil{Instituut voor Sterrenkunde, K. U. Leuven, Celestijnenlaan 200B, B-3001
Heverlee, Belgium}

\author{Helmut Feuchtgruber}
\affil{Max-Planck Institut f\"ur extraterrestrische Physik, Postfach 1603, D--85740,
Garching, Germany}

\and

\author{Gary J. Melnick}
\affil{Harvard-Smithsonian Center for Astrophysics, 60 Garden Street, Cambridge,
MA 02138}

\altaffiltext{1}{also Astronomy Department, Cornell University, Ithaca, NY, USA}
\altaffiltext{2}{also BARCO Projection Systems, Noordlaan 5, B-8520 Kuurne, Belgium}
\altaffiltext{3}{Based on observations with ISO, an ESA project
with instruments funded by ESA Member States (especially the PI countries: France, Germany,
the Netherlands and the United Kingdom) with the participation of ISAS and NASA.} 

\begin{abstract}
We combine spectra of VY CMa obtained with the short- and long-wavelength spectrometers,
SWS and LWS, on the Infrared Space Observatory to provide a first detailed continuum
spectrum of this highly luminous star.  The circumstellar dust cloud through which the star is
observed is partially self-absorbing, which makes for
complex computational modeling.  We review previous work and comment on the range of
uncertainties about the physical traits and mineralogical composition of the modeled disk.  We
show that these uncertainties significantly affect the modeling of the outflow and the estimated
mass loss. In particular, we demonstrate that a variety of quite diverse models can produce good
fits to the observed spectrum.  If the outflow is steady, and the radiative repulsion on the dust
cloud dominates the star's gravitational attraction, we show that the total dust mass-loss rate is
$\sim 4\times 10^{-6}M_{\odot}$\,yr$^{-1}$, assuming that the star is at a distance of 1.5 kpc. 
Several indications, however, suggest that the outflow from the star may be spasmodic.  We
discuss this and other problems facing the construction of a physically coherent model of the dust
cloud and a realistic mass-loss analysis.

\end{abstract}


\section{Introduction}

VY CMa is a highly luminous, variable M2/3 II supergiant. The star's spectrum is dominated by a
huge infrared excess. Le Sidaner \& Le Bertre (1996) have shown that the star is obscured,
with  $A_J$ = 3.09 - 3.35, corresponding to reddening E(B-V) $\sim$ 4.6. Only 1 percent of the
total luminosity is detected at optical wavelengths.  The star's luminosity has been estimated 
by a number of groups (cf Lada \& Reid, 1978, Le Sidaner \& Le Bertre, 1996,  Richards 
{\it et al.}, 1998 ), but remains somewhat uncertain because its distance is not 
well-established.  The consensus seems to converge on a distance of 1500 pc and a luminosity 
$L = 5\times 10^5 L_{\odot}$.

The star exhibits strong outflows. Buhl {\it et al.} (1975) detected thermal emission in the
vibrational ground state (v = 0, J = 2 - 1) 86.84686 GHz SiO line, which Reid and Dickinson
(1976) interpreted in terms of a stellar radial velocity of $17.6 \pm 1.5$\,km\,s$^{-1}$ with
respect to the local standard of rest, and an expansion velocity of 36.7$\pm 2.0$\,km\,s$^{-1}$.  
Neufeld {\it et al.} (1999) studied three thermally emitted mid-infrared water vapor emission
lines in a high-resolution spectrum obtained with the Short Wavelength Spectrometer (SWS) on
the Infrared Space Observatory (ISO), and found a mean radial velocity of order $20 \pm
2$\,km\,s$^{-1}$ and a $25$\,km\,s$^{-1}$ outflow velocity.  This outflow velocity is
significantly lower than the
velocities cited by Reid \& Dickinson (1976) or the 32\,km\,s$^{-1}$ that Reid \& Muhleman
(1978)
reported for 1612 MHz OH maser outflow.  

Le Sidaner \& Le Bertre (1996), who adopted a radial velocity of 20.5\,km\,s$^{-1}$, a
galactocentric distance to the Sun of 8.5 kpc, and a rotational velocity about the Galactic 
center of 220\,km\,s$^{-1}$ derived a distance to VY CMa,  $D = 2100$\,pc.  Earlier work by 
Lada \& Reid (1978) associates VY CMa with a molecular cloud complex, one edge of which
appears to
be illuminated by the star.  Since the radial velocities of the complex and star are rather 
similar, the star and cloud appear to lie at the same distance, which Lada \& Reid estimated at 
1500\,pc. On this basis they derived the star's luminosity as $L = 5\times 10^5 L_{\odot}$.  The 
difference in the distance estimates of Lada \& Reid on one hand, and Le Sidaner \& Le Bertre on 
the other, appears to be based on somewhat different modeling of the Galaxy, since both
estimates 
depend on roughly the same radial velocities.  

An independent means of estimating the star's distance is offered by the radial motions of
circumstellar masers.  VY CMa exhibits intense OH, H$_2$O and SiO maser emission indicating
rather complex spatial distributions and outflow velocities, particularly in the H$_2$O maser 
lines (Bowers, Claussen, \& Johnston, 1993).   Radio interferometric H$_2$O maser
observations that Richards {\it et al.} (1998) undertook after the pioneering observations by
Bowers, Claussen and Johnston indicate substantial proper motions in the nine years from 1985
to 1994, generally directed away from the assumed position of the central star.  If the star lies at a 
distance of 1500 pc, these proper motions translate into velocities of order 8 km\,s$^{-1}$ at 
a distance of 75 mas from the star and 32 km\,s$^{-1}$ at 360 mas.  Such velocities are 
comparable to the radial velocities of the H$_2$O masers, which exhibit a maximum expansion 
velocity of 35 km\,s$^{-1}$ at 400 mas from the central position. 

The effective temperature for VY CMa appears to be T$_{\rm eff}= 2,800$\,K, agreed on,
among others, by Monnier {\it et al.\/} (1999a), who adopt a distance of 1.5 kpc, and Le Sidaner
\& Le Betre (1996), who preferred a distance of 2.1 kpc to the star.   The combination of
effective temperature and a luminosity of $5\times 10^5 L_{\odot}$ then fix the star's radius as
$R_*\sim 14$\,AU, though Monnier {\it et al.}(1999a) cite a lower value $\sim 2\times 10^5
L_{\odot}$, and a correspondingly lower stellar radius of 8.8 AU on the basis of their
near-infrared observations ranging out to $\sim 3 \mu\,$m.  

High spatial-resolution, near-infrared observations of VY CMa have become available in the last
few years. Danchi {\it et al.} (1994) used heterodyne interferometry at 11\,$\mu$m, at intervals
that covered both maximum and minimum luminosity phases of the star. For these respective
phases, they model their data in terms of photospheric radii of 9.5 and 11 mas and inner radii for
the dust shell of 50 and 40 mas.  The outer radius of 2.5 arcsec, remains the same for both
phases.  A photospheric radius of 10 mas would correspond to a stellar radius of 15 AU for a star
at a distance of 1500 pc, in rough agreement with the conclusions of Lada \& Reid. 

Wittkowski {\it et al.} (1998) obtained  measurements of the circumstellar envelope's angular
diameter at 0.8, 1.28 and 2.17\,$\mu$m, which showed a nonspherical disk-like structure with
respective dimensions $60\times 83$, $80\times 166$ and $138\times 205$\,mas.  The star itself
was not resolved in their observations.  Monnier {\it et al.} (1999a) succeeded in mapping the
rather irregular dust cloud surrounding the star, at wavelengths of 1.65, 2.26 and 3.08\,$\mu$m.
They established that the  line of sight optical depths to the star at the respective wavelengths 
are 1.86 $\pm$ 0.42, 0.85 $\pm$ 0.20 and 0.44 $\pm$ 0.11.

Efstathiou \& Rowan-Robinson (1990) derived a model for VY CMa, which explains the object's
major features.  They consider the central star to be embedded in a massive disk of dust and gas
flowing out from the star at constant velocity.  The disk flares out from 
the star at angles $\pm 45^{\circ}$ above and below the disk's central plane.  The inner edge of 
the disk at radial distance $r_1$ has dust at a temperature $T_1 = 1000\,$K.  The disk stretches
to a radial distance $r_2$ from the star, with $r_1/r_2 = 0.006$.  At angles $> 45^{\circ}$ off 
the plane, radiation emitted by the star skirts over and under the disk and escapes into space.
Efstathiou and Rowan-Robinson believe that the line of sight from Earth is inclined at about
43$^{\circ}$ to the disk's orbital plane, so that we are looking almost tangentially through the
outer layers of the disk.  Their model accounts for the low levels of visible radiation as well as
the relatively high reddening observed by Monnier {\it et al.}(1999a).  The optical light we
receive from
the star is partly due to scattering, and is largely attenuated through absorption by 
circumstellar dust. The  radius $r_1$ in the model of Efstathiou \& Rowan-Robinson should
roughly
correspond to the figure of 40 - 50 mas given by Danchi {\it et al.}, or $\sim 70$\,AU, while 
their value of $r_2$ could correspond to either the value cited by Danchi {\it et al.} $\sim 
2.5$\,arcsec, or to the extent found by Monnier {\it et al.}(1999a) at 1.25$\,\mu$m, where the
cloud 
extends beyond 4 arcsec along some directions. These observations respectively yield  
$r_1/r_2\sim 0.02$ to 0.01, only slightly higher than the value suggested by Efstathiou \& 
Rowan-Robinson.

It is possible that the work of Efstathiou \& Rowan-Robinson could now be extended using
methods developed by Chiang and Goldreich (1997, 1999) for calculating the spectral energy
distribution of disks surrounding T Tauri stars.  While these stars are quite different from VY
CMa, and the disk surrounding VY CMa involves outflow rather than infall, several of the
insights developed by Chiang and Goldreich apply to both kinds of dust shells.

Mass loss rates cited for VY CMa by different authors range from $1\times
10^{-4}M_{\odot}$\,yr$^{-1}$ (Jura \& Kleinmann, 1990) to $3.1\times
10^{-4}M_{\odot}$\,yr$^{-1}$ (Danchi {\it et al.}, 1994), with other workers citing somewhat
different numbers.  All agree, however, that the mass loss rate is of this general order.

The structure of the present paper reflects the many significant uncertainties
remaining about VY CMa, in particular the wavelength-dependent optical depth of the
circumstellar cloud, the consequent acceleration and spatial density distribution of the outflow
from the star, and the chemical and mineralogical composition of the dust grains.  In sections 2
and 3 we describe, respectively,  the observations we carried out and the spectrum obtained.
Section 4 discusses the current state of modeling and the respective weaknesses of a broad range
of models that may provide excellent fits to the data but remain unphysical.  Section 5 introduces
an unabashedly unphysical toy model of VY CMa, primarily for the purposes of examining the
possible mineralogical constituents that might be expected in a satisfactory model of the VY
CMa outflow.  Section 6 explicitly lists the physical weaknesses of this model, but indicates why
the chemical nature of the various constituents might nevertheless be correctly identified. 
Sections 7 and 8 provide two different grain mass-loss rates from the star.  Sections 9 calls
attention to a minor anomalous spectral feature at $\sim 7\,\mu$m, for which we found no
satisfactory explanation.  A summarizing discussion and our conclusions follow, respectively, in
sections 10 and 11.

\section{Observations}

The observations on VY CMa reported here were obtained using both the short- and
long-wavelength spectrometers, SWS and LWS, on the Infrared Space Observatory, ISO (Kessler
{\it et al.} 1996). Because several hours of observing time were required to cover the entire
spectral range and because of scheduling conflicts the data were obtained on different ISO
orbits.  The wavelength ranges from 2.95 to 5.30 and 12.0 to 19.5\,$\mu$m were scanned on 13
October, 1997.  All other portions of the spectrum were scanned on the orbits of 19, 21 and 24
November, 1997. The observations of November 21 duplicated and extended the 2.95 to
5.30\,$\mu$m observations of October 13 and were in good agreement.   The wavelength range
from 12 to 19.5\,$\mu$m was not revisited during the November observing runs, but our
combined spectrum shows no discontinuities, suggesting that the star's spectrum did not
appreciably change during this period. This supposition is given weight by the observations of
Forrest, McCarthy and Houck (1979) who observed VY CMa from the NASA Lear Jet.  They
found the 20\,$\mu$m emission of VY CMa to be essentially constant in measurements carried
out in January and November, 1976 and September,  1977. In particular, the last two
measurements, taken ten months apart, differed by only 4\%.  Our ISO 20\,$\mu$m flux of
$8\times 10^{-15}$ W\,cm$^{-2}\,\mu$m$^{-1}$ is indistinguishable from their cited
September 1977 value, taken twenty years earlier.  This does not mean that the flux from VY
CMa is steady.  Monnier, Geballe \& Danchi (1999b) report 8 - 13\,$\mu$m variations on the
order of $\pm 20\%$ over several years, but significant variations are not expected on a time
scale
of 6 weeks for this irregular or semiregular variable whose approximate period is 2200 days.

The short wavelength spectrometer uses different apertures in scanning different parts of the
spectrum.  In the wavelength range from 2.38 to 12\,$\mu$m, the instrument views a $14"\times
20"$ rectangular beam on the sky.  At 12 to 29.5\,$\mu$m the corresponding beam dimensions
are $14"\times 27"$, and at 29.5 to 45.2\,$\mu$m, the beam size increases to $20"\times 33"$.
These beams all were centered on coordinates $\alpha = 7^h22^m58.^s3,\   \delta =
-25^{\circ}46'03"$.  The spectral resolving power $\lambda/\delta\lambda$ ranged from $\sim
1000$ to $\sim 2000$ in different portions of the spectrum. The total observing time on the star
was 10,185\,s.

A group led by Michael Barlow observed VY CMa with the long wavelength spectrometer
(LWS) on the night of 20 November, 1997.  Their spectrum obtained with a circular $\sim 80"$
diameter
field of view covered the wavelength range from 45 to 200\,$\mu$m in a run lasting 1930\,s
(Barlow 1999).  We made use of these data obtained from the ISO archives.  

Despite the difference in the sizes of the fields of view, we believe that even the smallest of 
the apertures used at the shortest wavelengths will have captured most of the radiation from the 
dust cloud enveloping the star. Our smallest field of view extends well beyond the faintest
near-infrared regions mapped by Monnier {\it et al.} (1999a).  We do not expect even the coldest
portions of the dust cloud radiating in the mid- and far-infrared, to lie outside the fields of 
view of the ISO spectrometers.

The flux integrated over our entire spectrum corresponds to $\sim 6\times
10^{-13}$\,W\,cm$^{-2}$.  This may be compared to a total flux of $\sim 7.1\times
10^{-13}$\,W\,cm$^{-2}$ expected from a star with bolometric luminosity $5\times 10^5
L_{\odot}$ at a distance of 1.5 kpc.  We, therefore, adopt an infrared dust cloud luminosity of
$\sim 4\times 10^5 L_{\odot}$.  In the model of Efstathiou and Rowan-Robinson the
outflow around the star occupies a disk that extends $\sim 45^{\circ}$ above and below its
central plane.  Such a disk occupies a field of view of $\sim 9\,$sr as seen from the star. If the
star's luminosity is $5\times 10^5 L_{\odot}$, roughly three-quarters of the star's radiation will
be absorbed by the disk and re-emitted isotropically in the infrared, while the remainder will
escape unattenuated in directions surrounding the disk axis of symmetry.  The total observed
infrared flux is consistent with this model and somewhat lower than expected if the star were
fully encased in a dust shell.

\section{Results}

The reduction of the SWS grating spectrum, which ranges from 2.38 to 45\,$\mu$m, was carried
out with the SWS Interactive Analysis system using calibration files from version 7.0 of the ISO
pipeline software.  The LWS spectrum obtained from the ISO archives had been reduced with
LWS pipeline 7.  Figure 1 shows a composite of all available data.

Figure 2. is a plot of the short-wavelength data out to 45\,$\mu$m, which shows a number of
highly broadened features and also registers some of the strongest spectral lines. The line
spectrum between 29.5 and 45\,$\mu$m has already been discussed by Neufeld {\it et al.},
(1999).

\section{Weaknesses of Current Models}

Some of the first models of VY CMa were primarily descriptive.  Early spectra of VY CMa
already revealed strong silicate emission bands at 10 and 20\,$\mu$m.  These motivated Le
Sidaner \& Le Bertre (1996) to use the ``dirty silicate" model of Jones \& Merrill (1976) in
discussing their observations.  

The first computational model to deal with both the mineralogy and geometry of the
circumstellar disk was due to Efstathiou \& Rowan-Robinson (1990), who took optical
depth effects into account and obtained a largely satisfactory fit to the spectrum of VY CMa. 
These models, though by now widely accepted in general outline, nevertheless entailed many
simplifications and somewhat arbitrary parameters.  One simplification was a $R^{-2}$
dependence of the density of the disk on radial distance from the star.  The second was the
density dependence on the altitude angle, i.e. the
angle above or below the central symmetry plane that the disk subtends at the star.  Efstathiou
and Rowan-Robinson assumed this to follow a power law up to some maximum ``opening angle"
of the disk, $\Theta$, though they note that this choice was ``not based on any physical grounds
but ... introduced in order to investigate its effects on the observable properties of the system." 
The third was the assumption of spherically symmetric scattering by the grains. With these three
caveats in mind, Efstathiou and Rowan-Robinson then computed a whole range of models with
different opening angles as seen by an observer placed at different viewing angles -- the choice of
viewing angle being a fourth free parameter.   

By examining a large number of power-law fits and viewing angles, in this class of models, the
best model fit, obtained on the further assumption that only ``dirty silicate" grains were involved,
suggested a central star surrounded by a disk with opening angle $45^{\circ}$ above and below
the plane, with the terrestrial observer viewing the star at an almost glancing angle of
$43^{\circ}$ above the disk, so that the line of sight to the observer still cuts through an
appreciable amount of obscuration.  Radiation the star emits at angles greater than $45^{\circ}$
above or below the disk propagates unattenuated into circumstellar space.
 
An important limitation of the Efstathiou and Rowan-Robinson model is that it neglects the
coupling between opacity and dynamics.  The dust grains absorb the star's radiation most strongly
at small distances from the star, where the outflow accelerates and the density correspondingly
drops more rapidly than $R^{-2}$.  Only at large distances from the star, where the radiation
pressure is low does the outflow reach a constant terminal velocity.  But at these distances the
irradiation also has dropped, and the grains evidently will hardly radiate at all.  The assumption
of constant velocity, therefore, may reflect a quite unphysical simplification. 

That Efstathiou and Rowan-Robinson, nevertheless obtain a good fit to the observed spectrum
may be seen as a sign of ambiguity.  A good model fit to the spectrum does not necessarily imply
a physically correct model. 

A first attempt to reduce this ambiguity by providing a model which couples radiative transfer to
dynamics was published by Ivezi\'{c} \& Elitzur (1995). These authors take into account not
only the radiation pressure acting on the dust, but also the countering gravitational attraction of
the star.  Zubko and Elitzur (2000) have recently applied this model to the outflow of water and
dust from the star W Hydrae.   Their model, however, is spherically symmetric and would have
to be modified to deal with the disk-shaped outflow now widely assumed to characterize VY
CMa, or with the plume of material that Monnier {\it et al.} (1999a) suggest may be flowing out
of the star, spraying its surroundings like a rotating garden hose.  

The present authors are not aware of any computer model currently available, that has
sufficient generality to permit a full description of the star.  

\section{A Toy Model Outflow}

In the present paper, we resign ourselves to not being able to provide a model that convincingly
couples radiative transfer to momentum transfer in a realistic outflow of condensing grains, and
limit ourselves to the presentation of our improved spectrum and the pursuit of a somewhat
complementary descriptive approach.  We set up a toy model that has many of the
same physical shortcomings as others but may help to illuminate the mineralogy somewhat
better, given the much more detailed spectral energy distribution now available.  In doing this,
we set aside the inspired  geometric approach of Efstathiou and Rowan-Robinson (1990), and the
elegant dynamic approach of Ivezi\'{c} and Elitzur (1995), in what is clearly no more than an
interim sketch.  

Our toy model is spherically symmetric and optically thin to radiation re-emitted by the grains. 
Both these assumptions are easily challenged, but even these unrealistic assumptions can yield a
remarkably good fit to the spectrum, and provide a number of insights to be enumerated below. 
The assumptions made permit us to use a straightforward version of the radiative transfer code
MODUST, which incorporates information on the mineralogical composition of the dust and
permits the introduction of several different mineralogical components.

In Figures 1 and 2 we present the MODUST fit obtained for the spectrum of VY CMa.  To
compare
results to other workers in the field, we have based all of these models on an assumed stellar
luminosity of  $5 \times 10^5 L_{\odot}$ and a distance to VY CMa of 1.5 kpc, even though the
flux we observe corresponds to a luminosity roughly 20\% lower, for a star at this distance. This
is consistent with our assumption that, as in the disk model of Efstathiou and Rowan-Robinson,
20\% of the flux escapes into space unattenuated, while 80\% is absorbed and
re-emitted in the infrared for an infrared luminosity of $4\times 10^5 L_{\odot}$.

Our model assumes a $R^{-2}$-dependence of the number-density of grains on radial distance
from the star.  It invokes two major mineralogical constituents, but also illustrates the lesser
contributions due to a potential admixture of two minor components.  The major constituents are
metallic iron and amorphous silicates, the minor components are crystalline silicates. We
included iron mainly because silicates are poor absorbers at short wavelengths where VY CMa
exhibits strong absorption. The minor components show that crystalline silicates can at best
provide only small contributions to the overall spectrum.  

Our choice of constituents is motivated by condensation sequences applicable to an
oxygen-rich star.   If we assume that the dust is formed from a vapor of solar composition during
the early stages of the outflow, metallic iron first condenses at T $\sim$ 1460 K, followed by
calcium-rich pyroxenes and nearly pure magnesium olivines at T $\sim$ 1450 K (see also
Whittet 1992).  Nuth {\it et al.} (2000) argue that, in a thermodynamic equilibrium condensation
sequence, magnesium silicates will condense well before iron silicates become stable at around 
600 K.  This makes it highly unlikely that $\mathrm{Fe}$-containing silicates will form on time
scales as short as those considered here, since the solid-state reaction rates between
$\mathrm{Fe}$-metal and magnesium silicate grains are slow.   Even when a
$\mathrm{Mg}$-$\mathrm{Fe}$-$\mathrm{SiO}$-vapor is condensed, the presence of
$\mathrm{Fe}$-atoms in the resulting grains excludes the presence of $\mathrm{Mg}$ and
vice-versa (Rietmeijer {\it et al.}, 1999).   Given these findings, we assume that the amorphous
silicates are pure $\mathrm{Mg}$-rich olivines. 

Our best fit to the data is shown in Figs. 1 and 2. The grain parameters used in the model are
given in Table 1.  We have used the optical properties of amorphous silicates that were available
from laboratory research, and emphasize that most of the parameters were chosen primarily to
obtain the best possible fit.  The temperature of the metallic iron particles in our model ranges
from 630 to 110 K, dropping from the inner to the outer shell, corresponding to distances from
the
star indicated in the second column of Table 1.  In general, one might expect the formation of
FeO and FeS instead of Fe at the cited temperatures.  However, oxidation of the metal grains,
which would give rise to a strong emission peak at 23 $\mu$m, appears to be forestalled by an
earlier incorporation of most of the free oxygen in H$_2$O and other molecules. Similarly, the
presence of triolite ($\mathrm{FeS}$) would give rise to several strong peaks in the near- and
mid-infrared, which are not discerned in the spectrum. 

Compared to iron, silicate grains absorb starlight less efficiently and re-radiate more efficiently at
long wavelengths.  They can therefore condense closer in to the star and still have a lower
temperature.  As Figure 2 demonstrates, the potential contributions from crystalline forsterite 
($\mathrm{Mg_2SiO_4}$) and crystalline enstatite ($\mathrm{MgSiO_3}$) are quite
minor, and it is not clear that these components are actually present.   We have included their
spectra in our model primarily to indicate the wavelengths at which contributions from these and
similar dust species would be significant and the extent to which small admixtures of different
grain materials could affect the spectral fit. Because the mid-infrared spectrum of VY CMa
exhibits small deviations on top of a smoother continuum emission, the presence of 
crystalline silicate dust is not unlikely; however, we cannot identify the actual carriers.    In this
respect the dust disk around VY CMa is quite different from the disk surrounding the young star
HD 100546 which exhibits strong forsterite emission  (Waelkens {\it et al}., 1996).

Two limitations now need to be mentioned:  

1. If we adopt the Solar System elemental abundances of Anders \& Grevesse (1989), the number 
densities of iron, magnesium and silicon atoms should all be very nearly identical. The atomic
weight of iron is 56 atomic mass units, amu.  Rietmeijer {\it et al.}, (1999) find that condensates
from Fe-Mg-SiO-H$_2$-O$_2$ vapor form MgO-SiO$_2$ solids that can take on the form of
pure amorphous silica SiO$_2$ and also pure MgO, but tend to cluster around compounds like
MgSiO$_3$, with equal numbers of magnesium and silicon atoms.  For MgSiO$_3$, the
molecular
weight is 100 amu, meaning that there is one atom of silicon plus one of magnesium per 100 amu
of silicates.  The iron-to-silicate mass ratios of 55 to 45 \% that we find in the best-fitting Toy
Model with parameters listed in Table 1, would then lead to elemental ratios of Fe:Si:Mg = 2:1:1. 
At face value, the iron abundance, therefore, appears to be rather higher than expected though not
too surprising, given the crudeness of the toy model.  

The importance of including iron in our models needs to be stressed.  Iron grains should naturally
condense at high temperatures and absorb well at near infrared frequencies where the star emits
much of its radiation but pure silicates absorb poorly (cf. Jones and Merrill, 1976).   The choice
by Efstathiou and Rowan-Robinson to base their model on ``dirty silicates" may lead to similar
spectra, but since the mineralogical nature of the ``dirt" in these silicates comes from empirical
fits found by Jones and Merrill, it does not readily lend itself to calculations of mass loss rates, or
an understanding of the chemistry of the outflowing material.

2.  A somewhat suspicious feature of our model is the sharp rise in the spectrum at short
wavelengths.  Out to about $\sim 5\,\mu$m it mimics the Wien distribution of a blackbody
at temperature $\sim 400\,$K.  This suggests that VY CMa is hidden from direct view
by a dust photosphere, which need not surround the star entirely but at least prevails for rays
passing radially outward through the disk.  The existence of such a photosphere is consonant
with the high near-infrared optical depths measured by Monnier {\it et al.} (1999a), which
indicate
that most of the starlight is absorbed by dust along the line of sight.  If the dust-photospheric
portion of the spectrum is close to blackbody radiation, then the chemical properties of the
radiating dust become less important, and the sharp rise in the observed spectrum at short
wavelengths no longer needs to be attributed to any particular constituent.  However, iron grains
do provide a means for readily obtaining substantial optical depths at wavelengths emitted by the
star.

We examine this question in the next section.   

\section{Weaknesses of the Toy Model}

The similarity of the iron emission curve in Fig. 2 to blackbody emission indicates that the
optical depth of the iron content of the model needs to be re-examined.  Our model assumed
spherical iron grains of radius $a$, with a size distribution $n(a)=  n(a_0)(a_0/a)^{3.5}$ and a
radial distribution proportional to $R^{-2}$, where $R$ is the distance from the star.  This radial
distribution corresponds to radial outflow of a homogeneous cloud at constant velocity, and
implies that shells of equal thickness at any distance from the star contain equal mass.   Calling
the inner radius of the iron dust cloud $R_0$, we obtain a particle number-density distribution
function 
\begin{equation}
n(a,R) = n(a_0,R_0)(a_0/a)^{3.5}(R_0/R)^2
\end{equation}
This leads to a total iron dust mass 
\begin{equation}
M = \int_{a_{min}}^{a_{max}}\int_{R_0}^{R_{max}}\biggl [\frac{4\pi}{3}a^3\rho\biggr ]
n(a_0,R_0)(a_0/a)^{3.5}(R_0/R)^2 [\Omega R^2] da dR
\end{equation}
or 
\begin{equation}
M = \frac{8\pi}{3}n(a_0,R_0)a_0^{3.5}\rho R_0^2(a_{max}^{1/2} -
a_{min}^{1/2})(R_{max} - R_0)\Omega
\end{equation}
Here $\rho$ is the density of individual iron grains, and $\Omega$ is the solid angle the dust
cloud subtends at the star.

The fraction of the radiation  $F$ absorbed by these grains, i.e. the fraction of the star's light
intercepted in a radial transit through the cloud for the optically thin case is 
\begin{equation}
F = \int_{a_{min}}^{a_{max}}\int_{R_0}^{R_{max}}Q_{abs} [\pi a^2 ]
\frac{n(a_0,R_0)(a_0/a)^{3.5}(R_0/R)^2 \Omega R^2}{\Omega R^2}da dR 
\end{equation}
In the present case, $F$ can exceed unity at some wavelengths, indicating that optical depth
effects need to be taken into account.  We can rewrite this equation as  
\begin{equation}
F = \frac{16\pi^2}{\lambda} Im\biggl
[\frac{m^2 -1}{m^2+2}\biggr ] n(a_0,R_0)a_0^{3.5}R_0^2[a_{max}^{1/2} -
a_{min}^{1/2}][R_0^{-1} - R_{max}^{-1}]
\end{equation}
where the Mie theory value for the absorption efficiency can be taken as $Q_{abs}= 8\pi
a/\lambda Im[(m^2-1)/(m^2+2)]$, $\lambda$ is the wavelength, and $Im$ indicates taking the
imaginary part of the function in square brackets for a complex index of refraction $m$.  The
ratio of absorption to mass is then 
\begin{equation}
\frac{F}{M} = \frac{6\pi}{R_0 R_{max}\rho\Omega\lambda} Im\biggl [\frac{m^2-
1}{m^2+2}\biggr ]
\end{equation}
For the Toy model, we used $\rho \sim 8$\,g\,cm$^{-3}$, $R_0 = 1.8 \times 10^5 R_{\odot}=
1.26
\times
10^{16}\,$cm, $R_{max} = 4.5 \times 10^6 R_{\odot} = 3.15\times 10^{17}$\,cm.   The total
iron dust-mass in the model is $M = 9.4\times 10^{-3}M_{\odot} \sim 1.9\times 10^{31}$\,g.
Using the Efstathiou \& Rowan-Robinson (1990) configuration, the dust disk occupies a field of
view $\Omega \sim 9$\,sr as seen from the star.  This leads to 
\begin{equation}
F = \frac{12.5}{\lambda} Im\biggl [\frac{m^2-1}{m^2+2}\biggr ]
\end{equation}
where the wavelength $\lambda$ is measured in micrometers.  Only $\sim 7\%$ of the star's
radiation comes from wavelengths as short as 0.73\,$\mu$m, where $F$ has as high a value as
$\sim 2.5$, indicating virtually total absorption at these wavelengths in a thin shell closest to the
star. At $1\mu$m, the complex index of refraction is $m\sim 3 + 4i$ (Pollack {\it et al.}, 1994),
so that the imaginary term is $\sim 0.1$.  At wavelengths longer than $\sim 1\,\mu$m, i.e. for
roughly three quarters of the energy output of a star at temperature 2800\,K, the
imaginary term rapidly diminishes to values less than 0.1, and $F$ even more rapidly diminishes
to a value below 1.25.  By $\lambda \sim 1.5 \,\mu$m, roughly the point beyond which the star
still emits half its light, $F$ has dropped to $\sim 0.5$ and $F$ should become a fairly good
indicator of the fraction of the radiation absorbed.  Taken together, these figures indicate
that virtually all the starlight at wavelengths short of 1\,$\mu$m, and more than half the light
between 1 and 1.5\,$\mu$m is fully absorbed in this model.  

Since the number of grains in shells of equal thickness around the star remains constant, and the
outflow velocity is taken to be constant, the differential absorption $F(R)$ declines with
increasing distance from the star as $R^{-2}$.  This has two immediate consequences for a cloud
that stretches radially from a distance $R_0$ out to much larger distances --- in our model out to
25 $R_0$.  Three quarters of whatever energy the cloud absorbs, is already absorbed by the time
radiation reaches a distance $2R_0$, so that the inner parts of the cloud closest to the star
act as though there were a relatively thin dust photosphere.  And since the absorption
integrated over most of the star's spectral energy distribution function over this radial
distance interval from the star is less than unity, the MODUST program will at least give roughly
reasonable results.  

The upshot of this argument is that the type of dust-photosphere which emerges from the model 
easily explains the steep blackbody-like rise in the VY CMa spectrum observed around
$4\,\mu$m, while also providing a superior fit at long wavelengths where a blackbody
photosphere would clearly give too low a flux. 

For the spectrum due to amorphous silicates, Draine \& Lee (1984) plot $Q_{abs}$ values,
which
indicate a range of $Im[(m^2 - 1)/(m^2+2) <0.04$ for all grains smaller than $1\,\mu$, at
essentially all wavelengths spanning the star's spectral energy distribution.  For the assumed
silicate grain density of 3 g\,cm$^{-3}$ in our model, this implies that
\begin{equation}
F = \frac{22.5}{\lambda} Im\biggl [\frac{m^2-1}{m^2+2}\biggr ] < 1
\end{equation}
so that the absorption in the region between $R_0$ and $2R_0$, where most of the radiation is
absorbed is always lower than 0.5 for grains 1\,$\mu$m in diameter and smaller.  The MODUST
program, therefore, again can be expected to yield reasonable values.

One might, however, still wonder why the Toy model silicate emission should so closely
resemble
that of the observations.  The answer to this is that silicates absorb roughly equally well at
1.5\,$\mu$m, where the star's spectral energy distribution is centered, as at 10\,$\mu$m, where
the blackbody photosphere or the iron photosphere of our Toy model have their respective
spectral peaks.  The curves drawn by Draine \& Lee (1984) show that $Im [(m^2-1)/(m^2+2)]$
rises to a value of $\sim 0.5$ around $10\,\mu$m and is remarkably independent of grain size for
the range $a = 0.01$ to 1$\,\mu$m.  This means that the fraction of the radiation absorbed  $F
\propto Im [(m^2-1)/(m^2+2)]/ \lambda$, and the implied grain absorption, is roughly the same
for photospheric emission from a star at temperature 2800\,K as for emission from a dust
photosphere of comparable luminosity.  The MODUST program, therefore, fortuitously provides
roughly correct results either way.

\section {Toy Model Mass-Loss Estimate} 

Despite the limitations just enumerated, our model should be able to yield a rough estimate of the
dust-mass loss from VY CMa.  The total dust mass in the Toy model is of the order of $1.6\times
10^{-2} M_{\odot}$.  The model
reaches out to a distance of $\sim 1550 R_* = 3.3\times 10^{17}\,$cm, rather larger than the
optical and near-infrared extent of the cloud measured by Monnier {\it et al.} (1999a), who
placed
the outer edge of their observed dust disk at around 4 arsec, or $\sim 10^{17}$\,cm.  For an
outflow velocity of 25-30 km s$^{-1}$ our model disk would have to replenish itself every
$3500-4200$\,yr.  The minimum dust-mass-loss rate thus would become about $\sim 3.8-4.6
\times
10^{-6}M_{\odot}$\,yr$^{-1}$ if we assume that all of the disk replenishes itself this rapidly.
If the gas-to-dust ratio in the outflow is of order 100, the grain mass loss we derive translates into
an overall mass of loss of $\dot M\sim 4\times 10^{-4}M_{\odot}$\,yr$^{-1}$. 
Both values are a factor of roughly 4 higher than estimates of the mass-loss obtained by other
observers.

Sopka {\it et al.\/} (1985) derived a grain mass loss of $\sim 8\times
10^{-7}M_{\odot}$\,yr$^{-1}$.  This estimate was  based on two considerations: an assumed
dust opacity law that ranges all the way from the ultraviolet to submillimeter wavelengths and a
400\,$\mu$m flux of 10\,Jy -- which can be compared to our best estimate of 15\,Jy.  The
remaining difference in our estimates therefore amounts to a factor of order 3, which is
understandable in view of the appreciable differences in approaches and assumptions.

Danchi {\it et al.} (1994) estimated a gas mass loss of $3.1\times
10^{-4}M_{\odot}$\,yr$^{-1}$, though their results cannot be directly compared, because they
assumed a mixture of silicates and graphite, and a temperature at the inner radius of the dust shell
as high as 1536\,K. The dust temperature they use in interpreting their visibility curves appears to
give too high a temperature and would result in higher near-infrared emission than we detect.
They, further, assumed a gas-to-dust mass-ratio of $\sim 200$, meaning that their dust mass loss
would amount to only $\sim 1.5\times 10^{-6}M_{\odot}$\,yr$^{-1}$.  

Our Toy model estimated mass loss is also rather higher than the mass loss of $1.2\times 10^{-4}
M_{\odot}\, {\rm yr}^{-1}$ that Netzer \& Knapp (1987) inferred for an adopted distance of
1900 pc, using CO (J = 2 - 1) data gathered by Zuckerman \& Dyck (1986). Since this value is
based on CO, rather than dust outflow, a direct comparison is also rather difficult.

\section{An Alternative Approach to a Mass-Loss Estimate}

A totally independent mass-loss estimate, rather simpler than those based on the more detailed
models described above, can be obtained on two assumptions.  The first is that all of the
observed infrared emission is emitted by dust and gas flowing out of the star with identical
outflow velocities.  The second is that the outward directed radiation pressure on the gas and dust
appreciably exceeds the gravitational attraction to the star.  

These assumptions would not apply to re-emission from dust in a gravitationally
bound photosphere.  The assumptions then specifically refer to a steady-state outflow as in our
Toy model, but might not hold if part of an emitting dust photosphere could remain
gravitationally bound and possibly re-contract in a variable star like VY CMa.   

To calculate the mass loss under the stated assumptions, we let the infrared luminosity of the dust
cloud be $L_{IR}$.  Then the rate at which outward directed momentum is deposited by the star
into the circumstellar disk is $L_{IR}/c$.  Since the terminal
velocity of the gas is known to be of order $v_{\infty} = 25$\,km\,s$^{-1}$, the total mass
accelerated from some initial velocity $v_0 <v_{\infty}$ per unit time amounts to 
\begin{equation}
\dot M = \frac{L_{IR}}{c(v_{\infty}- v_0)} \sim \frac{4\times 10^5 L_{\odot}}{7.5\times
10^{16}\ {\rm (cm/s)}^{2}}\sim 3\times 10^{-4} M_{\odot} {\rm yr}^{-1}
\end{equation}

Under the stated assumptions the numerical value on the right  is a minimum mass loss.  If there
were a substantial
injection velocity  $v_0$, the mass loss would be higher.

A more complete analysis that takes gravitational attraction into account leads to a more complex
conclusion. Let the acceleration of the cloud take place between radial distances $R$ and
$R+\Delta R$ from the star, and the outflow injection velocity at $R$ be $v_0$.  The
acceleration then is 
\begin{equation}
M\ddot R = \frac{L_{IR}}{c} - \frac{M_*MG}{R^2}
\end{equation}
where $M_*$ is the mass of the star, and $M$ the mass of the accelerating cloud.   Let  $v \equiv
v_0 +\Delta v$ be the velocity reached at distance $R+\Delta R$.  We readily integrate this with
respect to time.    For $R \gg \Delta R$, we obtain
\begin{equation}
\frac{M}{\Delta R/[v_0 +(\Delta v/2)]} =  \frac{L_{IR}}{ c\Delta v} -
\frac{M_*MG}{R^2\Delta v}
\end{equation}
The denominator on the left side of the equation is roughly the time during which the cloud is
accelerated to velocity $v_0 +\Delta v$, so that the full expression on the left gives the
approximate mass loss rate.  The first term on the right corresponds to the expression on the right
of equation (9).  If the gravitational attraction given by the second term on the right is substantial,
the mass loss rate not surprisingly decreases. (For a rigorous discussion of this topic, see
Ivezi\'{c} \&Elizur, 1995.)  

Equation (11) tells us that the very simple estimate of outflow given by equation (9) can only
yield an upper limit to the mass loss.  For VY CMa this upper limit, however, appears to be only
a factor of 2 to 4 higher than values based on the arguments of Sopka {\it et al.} (1985), Danchi
{\it et al.} (1994), or  Netzer \& Knapp, (1987). 

\section{An Anomalous Spectral Feature at 6.7 to 7.5\,$ \mu$m}

While we have been able to model most of the continuum spectrum of VY CMa through
emission by warm dust, an anomalous feature at 6.7 to 7.5\,$\mu$m has defied ready explanation.
We checked whether water vapor fluorescence might contribute, but find that it most probably
does not. 

Crovisier \& Encrenaz (1983) and Crovisier (1984) investigated the possibility of H$_2$O
fluorescence in comets, and focused attention on a few optically-excited transitions that appear 
of special significance.  Gonz\'alez-Alfonso \& Cernicharo (1999a) considered such processes
also for circumstellar envelopes.

Radiative absorption can raise a water molecule from the ground state, $\nu_1 \nu_2 \nu_3 =
000$, to the states 001 and 010 which, for illumination by cool stars, dominate by approximately
an order of magnitude.  Other transitions are less favored because the star emits too few photons
at the higher required excitation frequencies or because the molecular Einstein B coefficients are
too low. 

The sequence of events that could lead to the emission of significant radiation at $\sim 7$
microns involves an initial excitation 000 $\rightarrow$ 001, and subsequent decay 001
$\rightarrow$ 010 followed by 010 $\rightarrow$ 000.  If this process were in effect, we would
expect to also detect an absorption feature due to the 000$\rightarrow 001$ transition at
$2.66\,\mu$m and an emission feature at $\sim 4.36\,\mu$m due to the 001$\rightarrow 010$
transition.  Neither of these is apparent.

A second effect described by Gonz\'alez-Alfonso \& Cernicharo (1999a,b) involves solely
excitation from the ground vibrational state 000 to 010 and return to the ground state.  In the
downward transition from 010, some of the molecules end up in relatively excited rotational
states, from which they subsequently decay through purely rotational emission. In the downward
rovibrational transition to a higher rotational state the energy of the emitted photons is reduced
and their radiation is shifted toward longer wavelengths. A typical spectrum will therefore show
an absorption dip shortward of 6\,$\mu$m, and an increase beyond 6\,$\mu$m. 
Gonz\'alez-Alfonso and Cernicharo (1999b) have modeled this effect for evolved stars, but their
computations show that it is always more modest and restricted to individual spectral lines rather
than to a superposition of faint lines so dense as to produce a strong continuum.  In addition, the
energy in the hump observed at 7\,$\mu$m far exceeds any absorption below 6\,$\mu$m, and
also stretches to appreciably longer wavelengths than the model of Gonz\'alez-Alfonso and
Cernicharo predicts.  We therefore conclude that this feature is not produced by water vapor.

Tsuji {\it et al.} (1998) have tentatively identified a broad emission feature they observed in 
the M2III giant $\beta$ Peg at $\sim 7.6\,\mu$m with SiO.  This molecule should also exhibit a
feature at 4.0\,$\mu$m.  We detect no corresponding structure at 4\,$\mu$m, and our 7\,$\mu$m
feature falls at shorter wavelengths.

Dust species such as the silicate diopside (e.g. $\mathrm{CaMgSi_2O_6}$, Koike \& Shibai
1998) and crystalline spinel ($\mathrm{MgAl_2O_4}$, Chiwara {\it et al.} 2000) exhibit a broad
band at 7 $\mu$m (for spinel, this is not the case for a synthetic sample, but is true in natural
white and pink spinel found in Burma and Sri Lanka). This band is accompanied by stronger
bands at longer wavelengths. Amorphous diopside, which also exhibits the broad silicate bands
at 10 and 18 $\mu$m due to Si-O stretching and O-Si-O bending, is a particularly promising
candidate. More detailed future laboratory analysis will permit testing these minerals as
possible carriers of the 7 $\mu$m band. 

To date, we find ourselves unable to explain the 7\,$\mu$m feature, though we have no doubt
that it is real. The mismatches at longer wavelengths, e.g. at 25 and 34 $\mu$m can result from 
a slightly different chemical composition, ellipsoidal shape, or temperature of potential
crystalline silicate components. 

\section{Discussion}

In fitting the spectrum of VY CMa and estimating the star's mass loss, we have assumed that the
radiating circumstellar dust cloud is fairly homogeneous.  However, the patchy emission mapped
by Monnier {\it et al.\/} (1999a) suggests the existence of a plume of material possibly ejected in
different directions as the star rotates.

Ejection of individual plumes or blobs is also suggested by the 22-GHz water maser data of
Richards {\it et al.} (1998) who found a linear relationship between radial distance from the star
and radial displacement over the nine year period  from 1985 to 1994. This relation persists from
75 mas out to 350 mas ($\sim 100$ to 500 AU) and covers the velocity range from $\sim 8$
to $\sim 32$ km\,s$^{-1}$.  Remarkably, there is no apparent correlation between the observed
line of sight velocity of the masers and their displacement from the star, but the linearity between
proper motion and displacement extrapolates back to an intercept at zero
distance from the star.  Most workers in the field attribute the acceleration of both the thermal
gas cloud and the masers to drag by dust grains propelled outward by radiation pressure from the
star. Richards {\it et al.} find that the shell of water masers is elongated in the same direction
as the infrared emission from the dust, and that the inner radius of maser emission appears to 
lie just outside the inner dust shell radius.  The dust cloud and water masers, therefore, seem 
to be intimately related.   However,  the radiation pressure drops off as the inverse square of 
the distance, whereas a linear velocity/distance relation, $\dot r = ar$, where $a$ is a constant,
requires an acceleration, $\ddot{r} = a\dot{r} = a^2r$, proportional to distance from the star.
Even an optically thick cloud whose surface density declines as it expands radially outward from
a star will not produce this high an acceleration as a function of distance. 

The linearity of maser velocity with displacement from the star can be most easily explained if
all of the clouds producing maser emission were ejected at one and the same epoch and some
were accelerated to higher velocities than others.  Moreover, since the terminal velocity of the
masers is comparable to the velocity spread Neufeld {\it et al.} (1999) observed in the thermal
gas component, this would imply that the general stellar wind may similarly consists of
individual blobs accelerated to a range of different velocities.   This again implies that the bulk of
the acceleration occurs where the radiation pressure is highest, near the inner radius of the dust 
cloud, in a dust shell that is optically thick at visual and near-infrared wavelengths but 
may vary in mass, i.e. surface density.  The radiation pressure then accelerates low-mass blobs 
more rapidly and to higher terminal velocities than higher-mass blobs. 

The implication of this view is that individual blobs ejected from the star are independently
accelerated. This adds to the uncertainties still accompanying all models.
 
If the lowest density blobs have optical thickness unity, $\tau \sim 1$, for photospheric 
radiation, and grains of density $\rho\sim 4$\,g\,cm$^{-3}$ and radius $a \sim 0.5\,\mu$m occur 
with a gas-to-dust ratio $m_g/m_d\sim 100$, the blobs' column densities $\sigma$ would be
$\sigma\sim\tau d \rho m_g/m_d \sim 4\times 10^{-2}$\,g\,cm$^{-2}$, where $d$ is the grain
diameter.  The radiation pressure at
a distance $R\sim 600\,$AU $\sim 40 R_*$ from the star is $L_*/4\pi R^2 c\sim 6\times
10^{-5}$\,dyn cm$^{-2}$.  This leads to an acceleration of $\sim 1.6\times
10^{-3}$\,cm\,s$^{-2}$, so that a velocity of 30\,km\,s$^{-1}$ is reached in $t\sim 2\times
10^9$\,s, or $\sim 60$ years.  The distance the blob covers in that time is $\sim 200$\,AU, which
is sufficiently short that expansion and change in optical depth are quite minimal.  The
acceleration of the blob then stops primarily because  material ejected from the star somewhat
later begins to cool, form grains, absorb light, and thus cuts off the radiation from material at
greater radial distance from the star.  

If a blob at distance $R = 600$\,AU from the star, with column density $\sigma \sim 4\times
10^{-2}$\,g\,cm$^{-2}$ is ejected at intervals of $t\sim 2\times 10^9$\,s and surrounds the star
in solid angle of roughly 9\,sr, as in the Efstathiou \& Rowan-Robinson model, the mass outflow
will amount to $5\times 10^{29}$\,g\,yr$^{-1}$ or $\sim 3\times
10^{-4}M_{\odot}$\,yr$^{-1}$, in agreement with our steady-state-outflow model estimate.

Once a blob is accelerated to its terminal velocity the cloud material continues to drift outward,
shaded by material that was ejected later.  This shading is consistent with the observation by
Monnier {\it et al.} (1999a) that the optical depth of the cloud is of order 2 at 1.65\,$\mu$m, and
presumably higher at shorter wavelengths where most of the star's radiation is emitted.  An
implicit consequence of the argument presented in sections 6 and 8 is that infrared radiation
comes only
from the accelerating portions of the outflow.  Portions of the cloud flowing outward at constant
velocity do not experience radiative acceleration and neither absorb starlight nor emit infrared
radiation. 

\section{Conclusions}

We have used a very simple model to obtain a rather good spectral fit to the infrared spectrum of
VY CMa.  This and the inferred mass loss may constitute a good first approximation despite the
many uncertainties about this unusual star, and the many weaknesses of our model.  

Major constituents of the dust disk surrounding VY CMa appear to be amorphous silicates,
probably with a significant contribution by iron.  A good
spectral match obtained to the near-infrared excess with metallic iron, Fe, may be fortuitous,
though it would not be surprising, since iron particles should be first to form when a gas with
solar abundance cools (see e.g. Whittet 1992).  From condensation experiments by Nuth {\it et
al.} (2000), it is apparent that amorphous magnesium silicates are also likely to condense in
gases rich in silicon, oxygen, and magnesium.  Small admixtures of crystalline enstatite and
forsterite cannot be ruled out.  At least a fraction of the grains appears to be several microns in
diameter to account for the strong emission in the submillimeter and millimeter band. 

Our estimated mass-loss rate of order $4\times 10^{-4}M_{\odot}\,$yr$^{-1}$ is somewhat
higher than values cited by other authors using rather different assumptions.  However, all of the
estimates agree that this stage of the star's evolution cannot last much longer than another ten to
twenty thousand years, an interval during which 5 to 10 solar masses would be shed.  Unless VY
CMa is exceptionally massive, this is as much mass as the star may be expected to lose.

As discussed in detail in section 4, even the most thoughtful models presented to date remain
quite unphysical.  To date, no model has been devised that meets most of the criteria, namely: (a)
that it couples the dynamics to the radiative transfer, (b) that it incorporates a realistic
condensation sequence based on laboratory studies, (c) that it makes use of chemical abundances
that are credible, and (d) that its geometry be plausible.  Most models fail on all except perhaps
one or two of these four criteria, and so we have to grope our way forward slowly.

We have outlined the enormous amount of work that will be required before the nature of the
circumstellar dust cloud around VY CMa is fully understood.  Better computational tools, able to
deal with a variety of chemically differentiated, optically thin and thick grain components and a
more complex cloud geometry will be required and would lead to more realistic physical models. 
As input to such models, the condensation sequence of different chemical and mineralogical
species in an oxygen-rich environment will need to be determined from laboratory studies. These
could serve as input to computer models that couple an irradiated cloud's dynamical behavior to
its  density and temperature of condensed material in spherical, disk shaped or jet geometries.

\section{Acknowledgments}

The authors would like to thank both an anonymous referee and the editor for our manuscript,
Steven Willner, for the many improvements they recommended.  One of us (MH) acknowledges
support provided by NASA grant NAG5-3347. He is indebted, respectively, to Jose Cernicharo
and Joseph A. Nuth III for making available copies of their work before publication, and for
instructive suggestions. KM and LD acknowledge the Fund for Scientific Research - Flanders
(FWO-Vlaanderen). 

\vfill\eject
\centerline{\bf References}
\vskip 0.1 true in 
{\hoffset 20pt
\parindent = -20pt

Anders, E. \& Grevesse, N. 1989, Geochimica et Cosmochimica Acta  53, 197

Barlow, M. 1999, IAU proceedings N$^{o}$ 191, Asymptotic Giant Branch Stars, eds. T. Le
Bertre, A. L\`ebre \& C. Waelkens, 353

Bowers, P.F., Claussen, M. J. \& Johnston, K. J. 1993, AJ 105, 284

Buhl, D., Snyder, L. E., Lovas, F. J. \& Jonson, D. R. 1975, ApJ 201, L29

Chiang, E.I. \& Goldreich, P. 1997 ApJ 490, 368

Chiang, E.I. \& Goldreich, P. 1999 ApJ 519, 279

Chihara, H., Koike, C., Sogawa, H. \& Tsuchiyama, A. 2000, ASP Conf. Ser., Disks, Planets and
Planetesimals, eds.  F. Garz\'on, C. Eiroa, D. de Winter \& T. J. Mahoney., in press

Crovisier, J. \& Encrenaz, Th. 1983, A\&A 126, 170

Crovisier, J. 1984, A\&A 130, 361

Danchi, W.C., Bester, M., Degiacomi, C.G., Greenhill, L.J. \& Townes, C.H. 1994, \aj 
107, 1469

Draine, B.T. \& Lee, H.M. 1984, ApJ 285. 89

Efstathiou, A. \& Rowan-Robinson, M. 1990, MNRAS 245, 275

Forrest, W. J., McCarthy, J. F. \& Houck, J. R. 1979, ApJ 233, 611

Gonz\'alez-Alfonso, E. \& Cernicharo, J. 1999a, ApJ 525, 845

Gonz\'alez-Alfonso, E. \& Cernicharo, J. 1999b, in {\it The Universe as Seen by ISO}, eds. Cox,
P. \& Kessler, M.F., ESA-SP427, 325

Ivezi\'c, \v{Z}. \& Elitzur, M. 1995, ApJ 445, 415

J\"ager, C., Mutschke, H., Begemann, B., Dorschner, J. \& Henning, T. 1994, \aap 292, 641

J\"ager, C., Molster, F.J., Dorschner, J., Henning, 
T., Mutschke, H. \& Waters, L.B.F.M. 1998, \aap  339, 904

Jones, T.W. \& Merrill, K.M., 1976, ApJ 209 509

Jura, M. \& Kleinmann, S.G. 1990, ApJS  73, 769

Kessler, M.F., Steinz, J.A., Anderegg, M.E., Clavel, J., Drechsel, G., Estaria, P., Faelker, J.,
Riedinger, J.R., Robson, A., Taylor, B.G. \& Xim\'enez de Ferr\'an, S. 1996, \aap 315, L27

Koike, C., Shibai, H. 1998, in 'Report No. 671', The Institute of Space and Astronautical Science,
Kanagawa, Japan

Lada, C.J. \& Reid, M.J. 1978, ApJ 219, 95

Le Sidaner, P. \& Le Bertre, T. 1996, \aap  314, 896 

Marshall, C.R., Leahy, D.A. \& Kwok, S. 1992, PASP 104, 397

Monnier, J.D., Tuthill, P.G., Lopez, B., Cruzalebes, P., Danchi, W.C. \& Haniff, C.A. 1999a,
ApJ 512, 351 

Monnier, J.D., Geballe, T.R. \& Danchi, W.C. 1999b, ApJ 521, 261

Mutschke, H., Begemann, B., Dorschner, J., Guertler, J., Gustafson, B., Henning, T. \&
Stognienko, R. 1998, \aap 333, 188 

Netzer, N \& Knapp, G.R. 1987, ApJ 323, 734

Neufeld, D.A., Feuchtgruber, H., Harwit, M. \& Melnick, G. 1999, ApJ 517, L147

Nuth, J.A., Rietmeijer, F.J.M., Hallenbeck, S.L., Withey, P.A., \& Ferguson, F. 2000, ASP Conf. 
Ser. Vol. 196, Thermal Emission Spectroscopy and Analysis of Dust, Disks, and Regoliths, eds.  
M.L. Sitko, A.L. Sprague \& D.K. Lynch, 313

Pollack, J.B., Hollenbach, D., Beckwith, S., Simonelli, D.P., Roush, T. \& Fong, W. 1994, ApJ
421, 615

Reid, M. J. \& Dickinson, D. F. 1976, ApJ 209, 505.

Reid, M. J. \& Muhleman, D. O. 1978, ApJ 220, 229

Richards, A.M.S., Yates, J.A. \& Cohen, R.J. 1998, MNRAS 299, 319

Rietmeijer, F.J.M., Nuth, J.A. \& Karner, J.M. 1999, ApJ 527, 395

Sopka, R.J., Hildebrand, R., Jaffe, D.T., Gatley, I., Roellig, T., Werner, M., Jura, M. \&
Zuckerman, B. 1985, ApJ 294, 242

Tsuji, T., Ohnaka, K., Aoki, W. \& Yamamura, I. 1998, Ap\&SS 255, 293

van der Veen, W.E.C.J., Omont, A., Habing, H.J. \& Matthews, H.E. 1995, \aap 295, 445

Whittet, D.C.B. 1992, in 'Dust in the Galactic Environment', eds R.J. Taylor, R.E.

Wittkowski, M., Langer, N. \& Weigelt, G. 1998, \aap 340, L39

Zubko, V. \& Elitzur, M. 2000, ApJ 544, L137

Zuckerman, B. \& Dyck, H.M. 1986, ApJ 310, 207

}

\vfill\eject
\centerline{\bf Figure Captions}
\vskip 0.1 true in 
{\hoffset -20pt
\parindent -20pt

Fig. 1-- Overview of spectral data on VY CMa, obtained with ISO with the short wavelength
spectrometer SWS and the long wavelength spectrometer LWS (solid jagged curve).  Shown as
triangles, are IRAS data at 12, 25, 60, and 100\,$\mu$m, and ground-based submillimeter and
millimeter data (Sopka {\it et al.} 1985; Marshall {\it et al.} 1992; van der Veen {\it et al.}
1995).  Our estimate of the star's reddened flux is shown in the lower smooth curve, while the toy
model best fit for stellar plus circumstellar emission is shown in the upper smooth  curve (see
text).

Fig. 2 -- Infrared Continuum Spectrum of VY CMa (0-50 $\mu$m).  The dash-dotted line gives
our toy model fit (see text).  The shaded regions provide the relative contributions by different
dust components. The designation `Cr.' stands for `crystalline'.

}

\vfill\eject
\begin{table}[t]
\begin{center}
\begin{tabular}{|c||c|c|c|c|}
\hline \multicolumn{5}{c}{\Large Toy Model {\tt MODUST}: Fit Parameters}\\
\multicolumn{5}{c} {Parameters for the fit shown in Figure 1 and 2, together with their
references}\\
\hline
\multicolumn{2}{|c}{Stellar parameters:}&\multicolumn{3}{l|}{$\bullet$ distance d = 1500
pc}\\
\multicolumn{2}{|c}{}&\multicolumn{3}{l|}{$\bullet$ radius R$_{*}$ = 3000 R$_{\odot}$ }\\
\multicolumn{2}{|c}{}&\multicolumn{3}{l|}{$\bullet$ effective temperature T$_{\rm eff}$ =
2800 K }\\
\multicolumn{2}{|c}{}&\multicolumn{3}{l|}{$\bullet$ luminosity L$_{*}$ = 5 $\cdot$
10$^{5}$ L$_{\odot}$ }\\
\hline 
\multicolumn{2}{|c}{General dust parameters:}&\multicolumn{3}{l|}{$\bullet$ density law:
N(R) = N$_{R,o}$ (R/R$_{*}$)$^{-2}$}\\
\multicolumn{2}{|c}{}&\multicolumn{3}{l|}{$\bullet$ radius law: n(a) = n$_{a,o}$
(a/a$_{\rm min}$)$^{-3.5}$}\\
\hline 
 & Metallic $\mathrm{Fe}$ & Am. Sil. & Cr. $\mathrm{Mg_2SiO_4}$ & Cr.
$\mathrm{MgSiO_3}$\\ 
\hline
Inner radius (R$_{*}$) & 60 & 25 & 60 & 20\\
Outer radius (R$_{*}$) & 1500 & 1550 & 1000 & 500\\
N$_{r,o}$ (\,g\,cm$^{-3}$) & 3 $\cdot$ 10$^{-20}$ & 9 $\cdot$ 10$^{-20}$  &  2.7 $\cdot$
10$^{-22}$& 4.5 $\cdot$ 10$^{-20}$ \\
Particle radius, a, ($\mu$m)  & 0.01 - 1& 0.01 - 5& 0.01 - 1& 0.01 - 1\\
Grain density (\,g\,cm$^{-3}$) & 7.87 & 3.10 & 3.33 & 2.80\\
Dust temperature (K) & 110 - 630 & 50 - 440 & 90 - 360 & 60 - 170 \\
Total dust mass (M$_{\odot}$) & 9.4 $\cdot$ 10$^{-3}$ & 6.6 $\cdot$ 10$^{-3}$& 5.4 $\cdot$
10$^{-5}$& 5 $\cdot$ 10$^{-4}$\\
Relative percentage & 55 \% & 41 \% & 1 \% & 3 \% \\ 
\hline
\multicolumn{5}{c}{Optical properties used:}\\
\hline
\multicolumn{5}{c}{Amorphous silicates: J\"ager {\it et al.} (1994) \& Mutschke {\it et al.}
(1998)}\\
\multicolumn{5}{c}{Crystalline silicates: J\"ager {\it et al.} (1998)}\\
\multicolumn{5}{c}{Metallic $\mathrm{Fe}$: Pollack {\it et al.} (1994)}\\
\hline
\multicolumn{5}{c}{Notes:}\\
\hline
\multicolumn{5}{c}{`Cr.' indicates crystalline dust, while `Am.' indicates amorphous dust.}\\
\multicolumn{5}{c}{The internal grain density is not a free parameter but the density of the
mineral }\\ 
\multicolumn{5}{c}{studied in the laboratory to derive the optical constants of the respective
species.}\\ 
\multicolumn{5}{c}{All other parameters were chosen to provide a best fit to the data.}\\
\hline
\end{tabular}
\end{center}
\normalsize
\end{table}



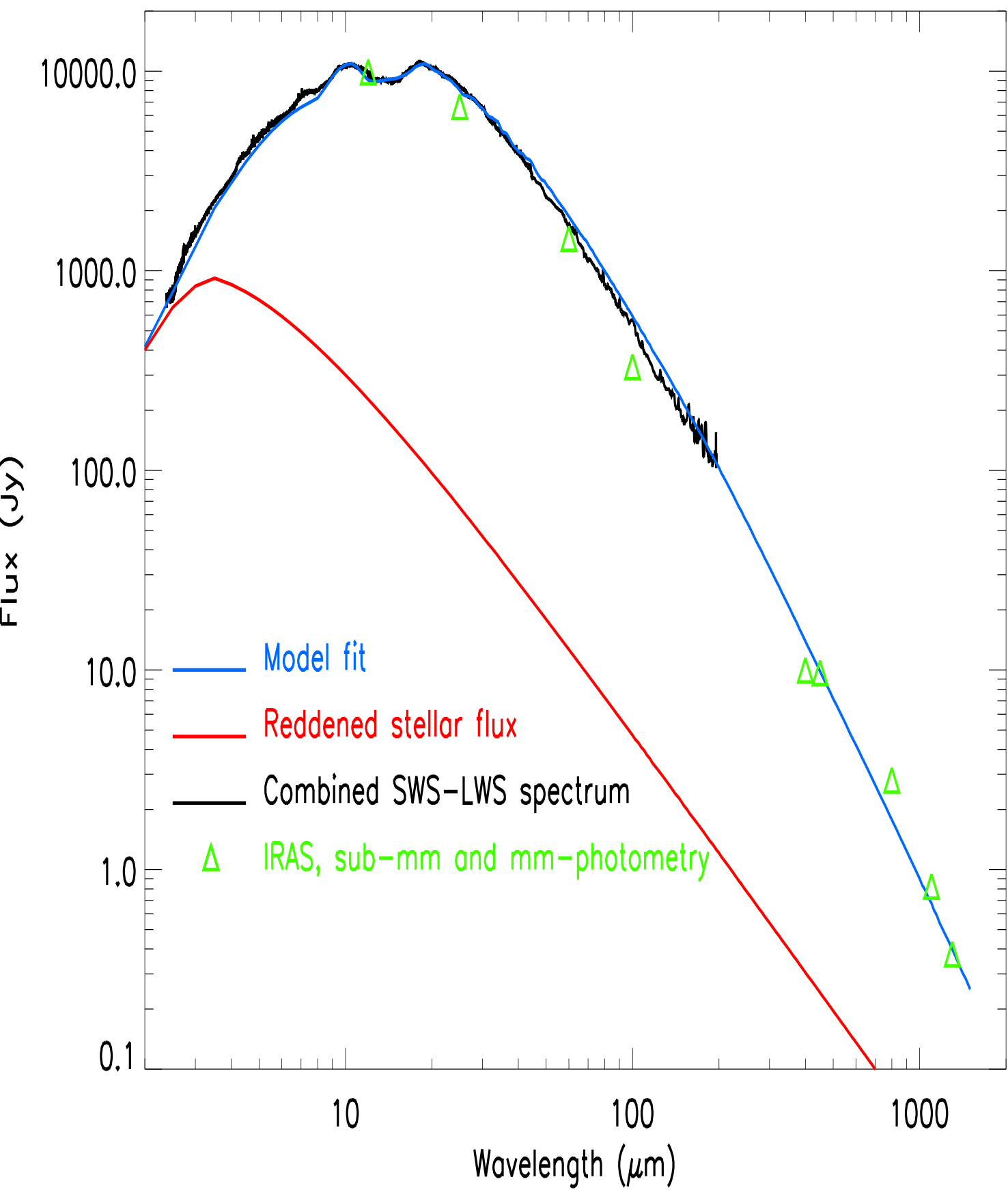
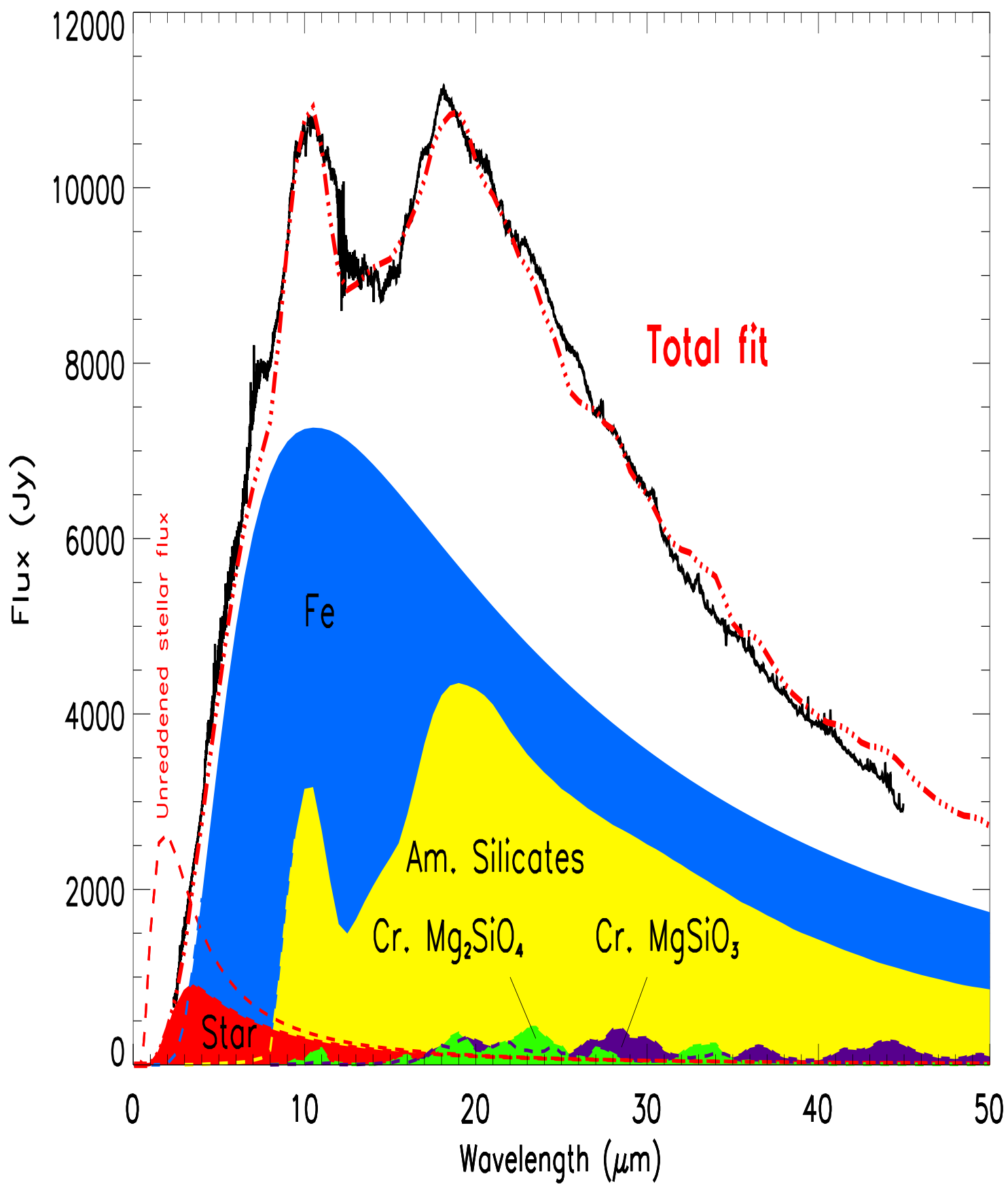
\end{document}